\documentclass[twocolumn,showpacs,preprintnumbers,amsmath,amssymb]{revtex4}
\usepackage{latexsym}
\usepackage{amsfonts}
\usepackage{amsmath}
\usepackage{bm}
\usepackage{amssymb}
\usepackage{subfigure}
\usepackage{epsfig}
\usepackage{graphicx}

\begin{document}

\title{Reducing multiphoton ionization in a linearly polarized
microwave field by local control}

\author{S. Huang$^1$}
\email{gtg098n@mail.gatech.edu}
\author{C. Chandre$^2$}
\author{T. Uzer$^1$}

\affiliation{$^1$ Center for Nonlinear Science, School of Physics,
Georgia Institute of Technology, Atlanta, Georgia 30332-0430, U.S.A.\\
$^2$ Centre de Physique Th\'eorique \footnote{Unit\'{e} Mixte de
Recherche (UMR 6207) du CNRS, et des universit\'{e}s Aix-Marseille
I, Aix-Marseille II et du Sud Toulon-Var. Laboratoire affili\'{e}
\`{a} la FRUMAM (FR 2291). Laboratoire de Recherche Conventionn¨¦
du CEA (DSM-06-35).} - CNRS, Luminy - Case 907, 13288 Marseille
cedex 09, France}

\date{\today}

\begin{abstract}
We present a control procedure to reduce the stochastic ionization
of hydrogen atom in a strong microwave field by adding to the
original Hamiltonian a comparatively small control term which
might consist of an additional set of microwave fields. This
modification restores select invariant tori in the dynamics and
prevents ionization. We demonstrate the procedure on the
one-dimensional model of microwave ionization.
\end{abstract}



\maketitle

\section{INTRODUCTION}
\label{sec1}

The multiphoton ionization of hydrogen Rydberg atoms
\cite{Gallagher9} in a strong microwave field \cite{Bayfield} is
an experiment which revolutionized the way we view the physics of
highly excited atoms \cite{Connerade} (for thorough reviews, see
\cite{Koch,Casati5,Casati,Jensen}). Its interpretation remained a
puzzle until its stochastic, diffusive nature was uncovered
through the then-new theory of chaos
\cite{Meerson,Casati5,Casati,Leopold3,Jensen1}. The multiphoton
ionization is believed to occur when the electrons diffuse to
increasingly higher energies chaotically by taking advantage of
the breakups of local invariant tori in phase space
\cite{Casati5,Casati,MacKay,Farrelly,Howard1,Howard2,Reichl}.
Classical theory can be applied to the ionization of hydrogen in
the parameter regime where the microwave and Kepler frequencies
are nearly equal \cite{Jensen,MacKay}. Because the classical
dynamics of this system is chaotic, the Rydberg states of hydrogen
are an excellent testbed for investigating the quantal
manifestations of classical chaos \cite{Blumel9,Buchleitner,Krug},
i.e., the field of ``quantum chaology"
\cite{Casati5,Casati,Reichl,Berry}. Indeed, the literature on the
correspondence between classical and quantum behavior in the
ionization of Rydberg atoms is extensive
\cite{Meerson,Casati5,Casati,Leopold3,Jensen1,MacKay,Farrelly,Howard1,Howard2,
Reichl,Blumel9,Buchleitner,Krug,Berry,Jensen,Leopold,Leopold1,Blumel,Delande,Perotti}.
We perform our purely classical calculations in the regime where
the quantum-classical correspondence is particularly close
\cite{Jensen}.

Recently, the research focus in this field has shifted from
understanding to manipulating the ionization process
\cite{Ko,Sirko1,Sirko2,PMKoch,Gallagher,Maeda9}. Since microwave
ionization of Rydberg states is a paradigm for time-dependent
nonintegrable systems, learning to manipulate stochastic
ionization is expected to pave the way to controlling other, more
involved systems. The control of stochastic ionization has been
investigated using both quantum and classical approaches in the
past few years \cite{Sirko2,PMKoch,Gallagher,Sirko1,Prosen}. Here,
we return to the basic dynamics of the stochastic ionization
process to answer the most elementary manipulation question: If
the multiphoton ionization is made possible by broken invariant
tori, can ionization be reduced (or even stopped) by restoring
invariant tori at carefully chosen locations in phase space?

In this paper, we will show how to reduce or shut off the
ionization of Rydberg atoms using a ``local" control strategy
which originates in plasma physics \cite{Chandre2}. The premise of
the procedure is to reduce the chaos (and thus ionization) in a
selected parameter range through a small perturbation which
regularizes the dynamics in that narrow area but does not affect
the dynamics elsewhere. Technically, local control achieves this
by creating an invariant torus in a selected region of phase space
without significantly changing other parts of phase space.

The problem of finding such a modification of the original
Hamiltonian system is, a priori, nontrivial~: A generic
modification term would lead to the enhancement of the chaotic
behavior (following the intuition given by Chirikov's criterion
\cite{Chirikov}). Modification terms with a regularizing effect
are, of course, rare. However, there is a general strategy and an
explicit algorithm to design such modifications which indeed
drastically reduce chaos and its attendant diffusion by building
barriers in phase space \cite{Ciraolo1,Chandre2}, as we will show
on the one-dimensional hydrogen Rydberg atom in a microwave field.

One-dimensional models of microwave ionization in linearly
polarized microwave fields have proven perfectly adequate to
explain most experimental observations
\cite{Koch,Casati5,Casati,Jensen} since many of the experiments
considered extended, quasi-one-dimensional hydrogen atoms
\cite{Bayfield,Koch,Reichl,Jensen} in which the angular momentum
of the Rydberg electrons is much smaller than their principal
quantum number. As a result, the atoms resemble needles in which
the electron bombards the core with zero angular momentum. The
Hamiltonian in atomic units reads

\begin{equation}
\label{Hatom} H(p,x,t)=\frac{p^{2}}{2}-\frac{1}{x}+\lambda
x\cos\omega t,
\end{equation}
where $\lambda$ is the amplitude of the external field.

The desired Hamiltonian with the control field reads
\begin{equation}
\label{Hc} H(p,x,t)=\frac{p^{2}}{2}-\frac{1}{x}+\lambda x\cos
\omega t+x f(t).
\end{equation}
For practical purposes, we expect that the control field $f(t)$
has the same form as the perturbation field but with relatively
small amplitude. Despite the fact that $f$ introduces an
additional set of resonances, its effect, if it is appropriately
chosen, is to restore specific invariant tori.

This paper is organized as follows: In Sec.~\ref{sec2}, after
summarizing the control method \cite{Chandre2}, we implement it on
a one-dimensional hydrogen atom driven by a linearly polarized
microwave field. In Sec.~\ref{sec3}, we present the numerics of
the control term and show its efficiency by using Poincar\'e
sections, laminar plots and diffusion curves. In order to be
relevant and feasible for physical implementations, the control
term has to be robust, i.e., sufficiently good approximations to
it should reduce chaos effectively, too. We pay particular
attention to this point and show numerically that reasonable
approximations to our control terms are effective in reducing
chaos also. Conclusions are in Sec.~\ref{sec4}.

\section{Computation of the control term}
\label{sec2}

We consider Hamiltonian systems with $L$ degrees of freedom,
written into action-angle variables $({\bf A},{\bm\theta})$ of the
form
\begin{equation}
H({\bf A},{{\bm\theta}})=H_{0}({\bf A})+\varepsilon V({\bf
A},{\bm\theta}),
\end{equation}
where $({\bf A},{\bm\theta})\in {\mathbb R}^{L}\times {\mathbb
T}^{L}$.

In the integrable case ($\varepsilon=0$), the phase space is
foliated by invariant tori with frequency ${\bm\omega}({\bf
A})=\partial H_0/\partial{\bf A}$. Let us consider one of these
invariant tori with frequency $\bm{\omega}$ and position ${\bf
A}_0$. We assume that $\bm\omega$ is non-resonant, i.e.\ there is
no non-zero integer vector $\bf k$ such that ${\bm\omega}\cdot
{\bf k}=0$. This invariant torus is generally destroyed by the
perturbation $V({\bf A},{\bm\theta})$ when the parameter
$\varepsilon$ is greater than a critical value $\varepsilon_{c}$.
The idea of the control is to rebuild this invariant torus with
frequency $\bm{\omega}$ for $\varepsilon>\varepsilon_{c}$ by
adding a small control term $f$ to the original Hamiltonian $H$.
Thus the controlled Hamiltonian $H_c$ can be constructed as
$$
H_{c}({\bf A},{\bm \theta})=H({\bf A},{\bm \theta})+f({\bm
\theta}).
$$
The expression of control term is given by
\begin{equation}
\label{eqnfth} f(\bm{\theta})=-H({\bf
A}_{0}-\partial_{\bm{\theta}}\Gamma b(\bm{\theta}),\bm{\theta}),
\end{equation}
where $b(\bm{\theta})=H({\bf A}_{0},\bm{\theta})$ and $\Gamma$ is
a linear operator defined as a pseudo-inverse of $\bm{\omega}\cdot
\partial_{\bm{\theta}}$. Its explicit expression is
\begin{equation}
\Gamma b(\bm{\theta})=\sum_{\bm{\omega}\cdot \bm{k}\neq
0}\frac{b_{\bm{k}}}{i\bm{\omega}\cdot
\bm{k}}e^{i\bf{k}\cdot\bm{\theta} }.
\end{equation}
for $b(\bm{\theta})=\sum_{\bm{k}\in {\mathbb
Z}^{L}}b_{\bm{k}}e^{i\bf{k}\cdot\bm{\theta}}$. The restored
invariant torus of the controlled Hamiltonian $H_{c}$ has the
equation~:
\begin{equation}
{\bf A}={\bf A}_{0}-\Gamma\partial_{\bm{\theta}}H({\bf
A}_{0},\bm{\theta}).
\end{equation}
Such an invariant torus acts as a barrier to diffusion for
Hamiltonian systems with two degrees of freedom.

In order to apply this method to a one-dimensional hydrogen atom
driven by a microwave field, we first need to map
Hamiltonian~(\ref{Hatom}) into action-angle variables of the
unperturbed system $(\lambda=0)$. Its action-angle variables
$(J,\theta)$ are \cite{Leopold}
\begin{eqnarray*}
&&x=2J^{2}\sin^{2}\varphi,\\
&&p=\frac{1}{J}\cot\varphi,
\end{eqnarray*}
with
$$
\theta=2\varphi-\sin2\varphi.
$$
After rescaling energy, time, position and momentum as
$H'=\omega^{-2/3} H, t'=\omega t, x'=\omega^{2/3}x,
p'=\omega^{-1/3}p$, we obtain the rescaled field amplitude
$\lambda'=\omega^{-4/3}\lambda$. The rescaled Hamiltonian still
satisfies the equations of motion, and we assume $\omega=1$
without loss of generality in Eq.~(\ref{Hatom}). The scaled
frequency, or say, winding ratio in the rescaled system is thus
defined as $\varpi\equiv J^3\omega=J^3$. Expanding $x'$, we
rewrite Hamiltonian~(\ref{Hatom}) \cite{Casati5}
\begin{equation}
\label{HatomAA}
H=-\frac{1}{2J^{2}}+2J^{2}\lambda(\frac{a_{0}}{2}+\sum_{n=1}^\infty{a_{n}\cos
n\theta})\cos t,
\end{equation}
where
$$
a_{n}=\frac{J_{n}(n)-J_{n-1}(n)}{n},
$$
and $J_{n}$'s are Bessel functions of the first kind. We
abbreviate the Hamiltonian~(\ref{HatomAA}) as
$$
H=-\frac{1}{2J^{2}}+\lambda J^{2}v(\theta,t).
$$
where
\begin{equation}
\label{eqnv} v(\theta,t)=a_{0}\cos t
+\sum_{n=1}^\infty{a_{n}[\cos(n\theta+ t)+\cos(n\theta- t)]}.
\end{equation}
The Hamiltonian~(\ref{HatomAA}) displays a set of primary
resonances approximately located at $J_{n}=n^{1/3}$. The overlap
of these resonances~\cite{Chirikov} leads to large-scale chaos and
hence ionization. We expect the lower action region to be more
regular, and it is chaotic for sufficiently large $\lambda$ by
resonance overlap. The idea is, given a value of $n$, to restore
an invariant torus in between the resonances approximately located
at $J_n$ and $J_{n+1}$. For $\lambda=0$, this invariant torus with
(Kepler) frequency $\omega_0$ is located at
$J_0=\omega_{0}^{-1/3}$. In order to do this, we compute the
control terms as explained above.

The next step of the control algorithm is to map the
time-dependent Hamiltonian into an autonomous one. We consider
that $t$ (modulus $2\pi$) is an additional angle variable and we
call its corresponding action variable $E$. The autonomous
Hamiltonian becomes $H(J,\theta,t)+E$. The action-angle variables
are ${\bf A}=(J,E)$ and ${\bm\theta}=(\theta,t)$.

The frequency vector of the torus is $\bm{\omega}=(\omega_{0},1)$.
The formula of the control term is obtained by replacing the
actions $\bf A$ by ${\bf A}_0-\partial_{\bm\theta}\Gamma H({\bf
A}_0,{\bm\theta})$ where ${\bf A}_0=(J_0,0)$. The control term $f$
is given by
\begin{eqnarray}
f(\theta,t)&=&\sum_{k=2}^\infty{\frac{k+1}{2}\lambda^{k}\omega_{0}^{\frac{2-k}{3}}(\Gamma\partial_\theta v)^{k}}{}\nonumber\\
&&{}+(2\lambda^{2}\omega_{0}^{-1}\Gamma\partial_\theta
v-\lambda^{3}\omega_{0}^{-\frac{4}{3}}(\Gamma\partial_\theta
v)^{2})v,\label{eqnf}
\end{eqnarray}
where
$$
\Gamma\partial_\theta
v=\sum_{n=1}^\infty{na_{n}}\left[\frac{\cos(n\theta+
t)}{n\omega_{0}+1}+\frac{\cos(n\theta- t)}{n\omega_{0}-1}\right].
$$
For $\lambda$ small, we approximate the control term $f$ by its
leading order in $\lambda^2$ which is given by
\begin{equation}
\label{eqnf2}
f_2(\theta,t)=\frac{3}{2}\lambda^{2}(\Gamma\partial_\theta
v)^{2}+2\lambda^{2}\omega_{0}^{-1}v\Gamma\partial_\theta v.
\end{equation}

Obviously, the location of the restored invariant torus depends on
the choice of Kepler frequency $\omega_{0}$ or equivalently of its
location in the integrable case $J_0$. The theoretical torus curve
is given by
\begin{equation}
J=J(\theta,t)=\omega_{0}^{-\frac{1}{3}}-\lambda
\omega_{0}^{-\frac{2}{3}}\Gamma\partial_\theta v(\theta,t).
\end{equation}
This torus is $\lambda$-close to $J_0=\omega_{0}^{-1/3}$. We
notice that the control term as well as the invariant torus are
$2\pi$-periodic in $\theta$ and time $t$.

{\em Remark}~: In the local control method, we have searched for
control terms only dependent on $\bm \theta$. However, in order to
be more consistent with the specific shape of the control waves,
it can be appropriate to search for controlled Hamiltonian of the
form
$$
H_c({\bf A},{\bm\theta})=H({\bf
A},{\bm\theta})+({\bm\Omega}\cdot{\bf A})^2 f({\bm\theta}),
$$
where $\bm\Omega$ is a fixed vector, e.g., ${\bm\Omega}=(1,0)$ in
that case. Following the same arguments as in
Ref.~\cite{Chandre2}, the formula of the control term is
\begin{equation}
\label{eqnfoa} f({\bm\theta})=-\frac{H({\bf
A}_0-\partial_{\bm\theta}\Gamma
b,{\bm\theta})}{\left({\bm\Omega}\cdot{\bf A}_0- {\bm
\Omega}\cdot\partial_{\bm\theta}\Gamma b\right)^2},
\end{equation}
where $b({\bm\theta})=H({\bf A}_0,{\bm\theta})$. We notice that
the control term~(\ref{eqnfoa}) is still of the same order as the
one given by Eq.~(\ref{eqnf}) and it is $\varepsilon^3$-close to
the one given by Eq.~(\ref{eqnfth}) divided by
$({\bm\Omega}\cdot{\bf A}_0)^2$, since $\partial_{\bm \theta}b$ is
of order $\varepsilon$.

\section{Numerical analysis}
\label{sec3}

In what follows, the series which give $v$ and
$\Gamma\partial_\theta v$ are truncated at $n=30$ for numerical
purposes, and the first series of $f$ is truncated at $k=20$.
Also, we choose $\omega_{0}$ in the interval
$[\frac{1}{n_0+1},\frac{1}{n_0}]$ which corresponds to a region in
between two primary resonances, and $n_0$ is in general chosen
equal to 1, 2, 3... (With a relatively big $n_0$, local control
theory still holds though quantum suppression leads to a higher
ionization threshold \cite{Casati9}).

\subsection{Analysis of the control term}
\label{sec3A}

\begin{figure*}
 \begin{minipage}[t]{8cm}
 \centering
 \includegraphics[width=6.cm,height=5.4cm]{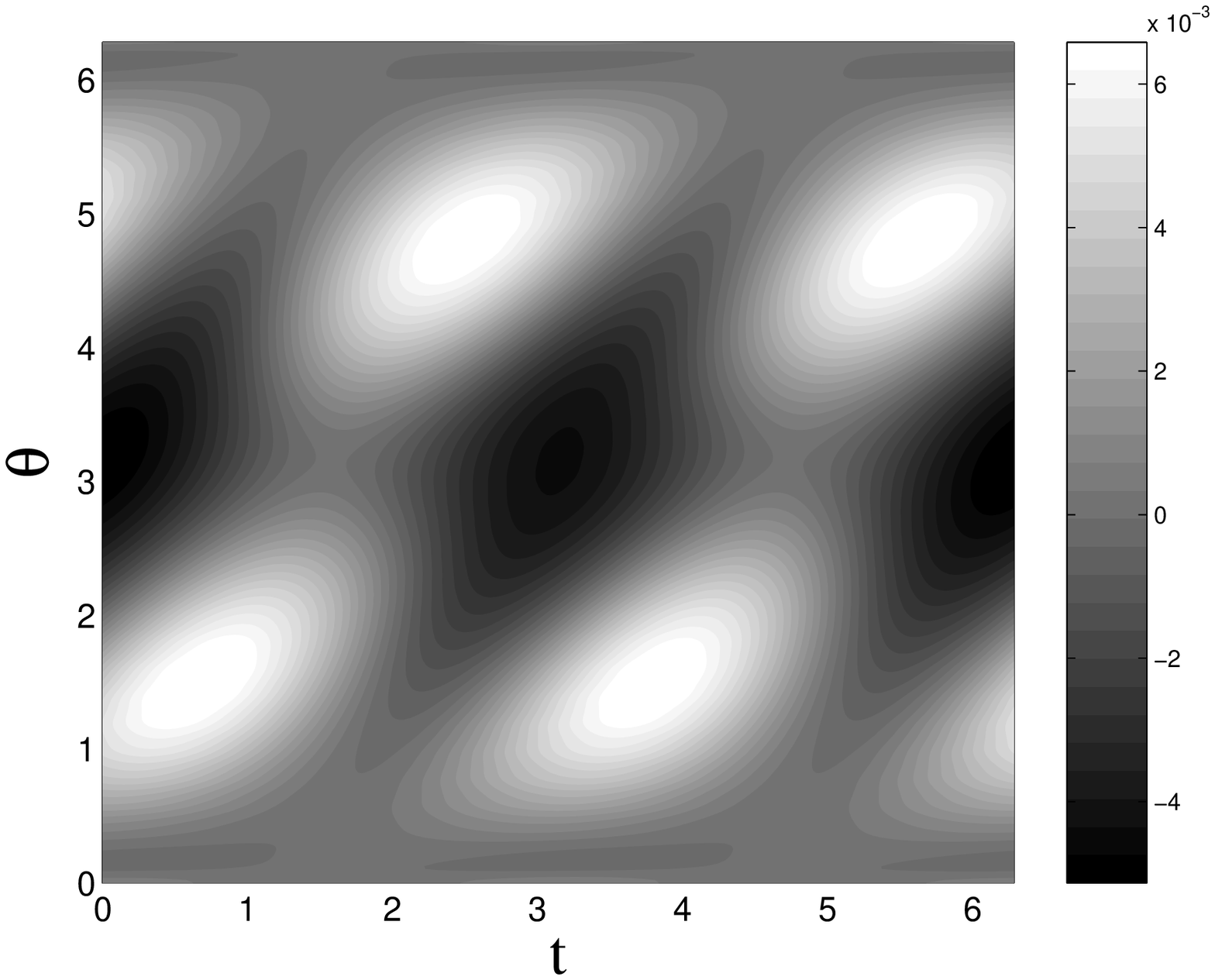}
 \mbox{{\bf (a)} }
 \end{minipage}
 \begin{minipage}[t]{8cm}
 \centering
 \includegraphics[width=6.cm,height=5.4cm]{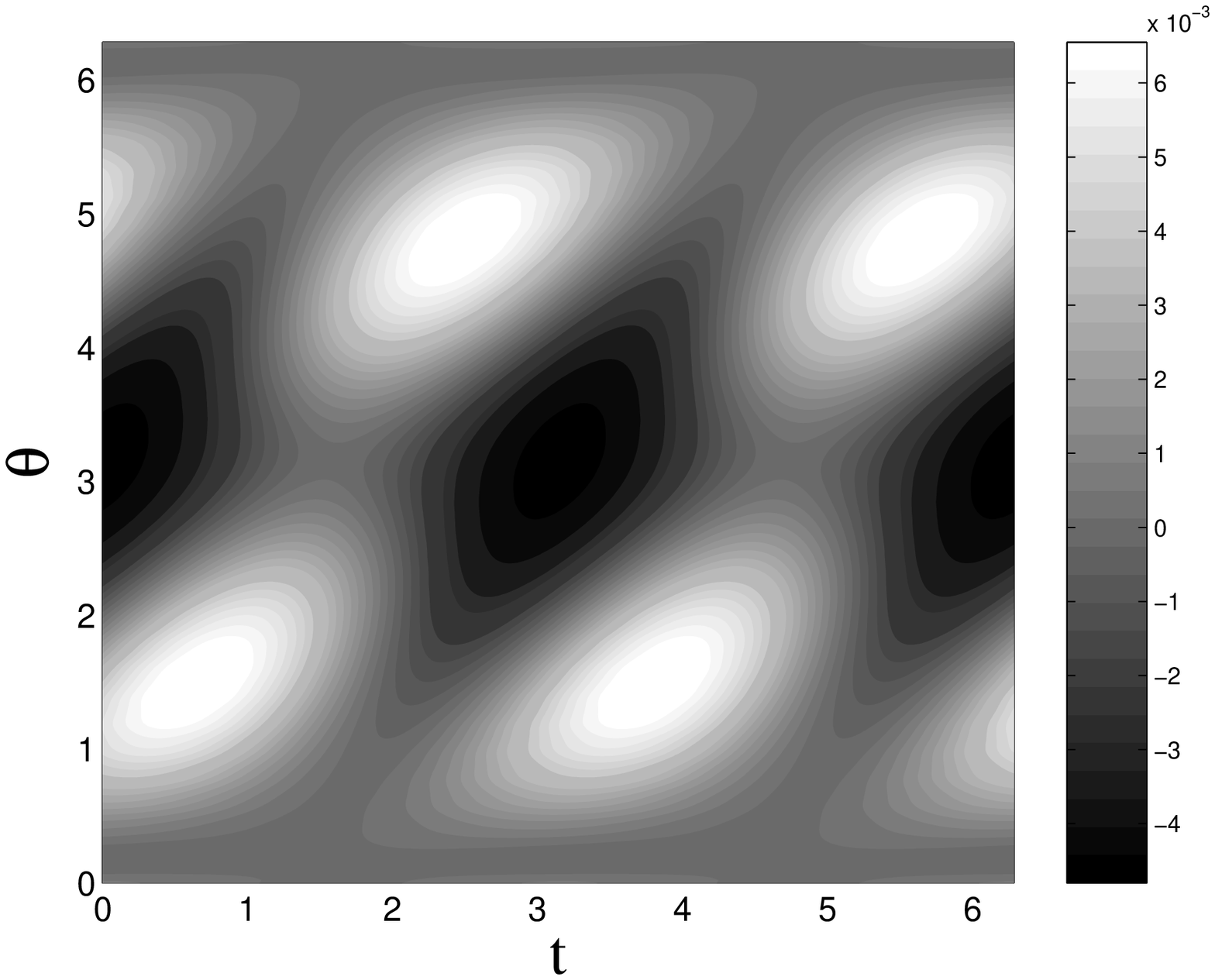}
 \mbox{{\bf (b)} }
 \end{minipage}
 \caption {\label{fig:fig1} Contour plots of (a) $f$ given by Eq.~(\ref{eqnf}) and (b) $f_2$ given by Eq.~(\ref{eqnf2})
 for $\lambda=0.03$ and $\omega_{0}=0.6750$.}
\end{figure*}

\begin{figure*}
 \centering
 \includegraphics[width=6.cm,height=5.4cm]{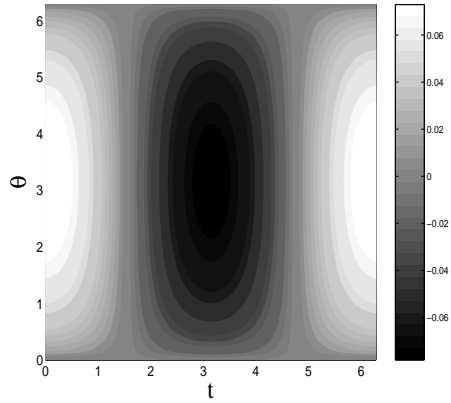}

 \caption{\label{fig:fig2} Contour plot of $\lambda J_{0}^{2}v(\theta,t)$ where $v$ is given by Eq.~(\ref{eqnv}) for $\lambda=0.03$ and $\omega_{0}=0.6750$.}
\end{figure*}

Figure~\ref{fig:fig1} depicts a contour plot of $f$ given by
Eq.~(\ref{eqnf}) and $f_2$ given by Eq.~(\ref{eqnf2}) for
$\omega_0=0.6750$ (which corresponds to $n_0=1$) and
$\lambda=0.03$. In this case, the scaled frequency at the intended
invariant torus for $\lambda=0$ is
$\varpi=J_{0}^3=\omega_{0}^{-1}=1.4815<2$ which justifies the
application of classical theory to the regime we are interested
\cite{MacKay,Galvez}. Since $f$ and $f_2$ are $2\pi$-periodic in
$t$ and $\theta$, these contour plots are represented for
$(t,\theta)\in [0,2\pi]^2$. In order to compare the control term
with the perturbation, Fig.~\ref{fig:fig2} represents a contour
plot of the perturbation at an action $J=J_0$ where the control
acts. These figures show that for this value of $\lambda$ the
control term is small (by a factor approximately equal to 10)
compared with the value of the external field $\lambda J_0^2
v(\theta,t)$.

\begin{figure*}[!htb]
 \begin{minipage}[t]{8cm}
 \centering
 \includegraphics[width=6.cm,height=5.4cm]{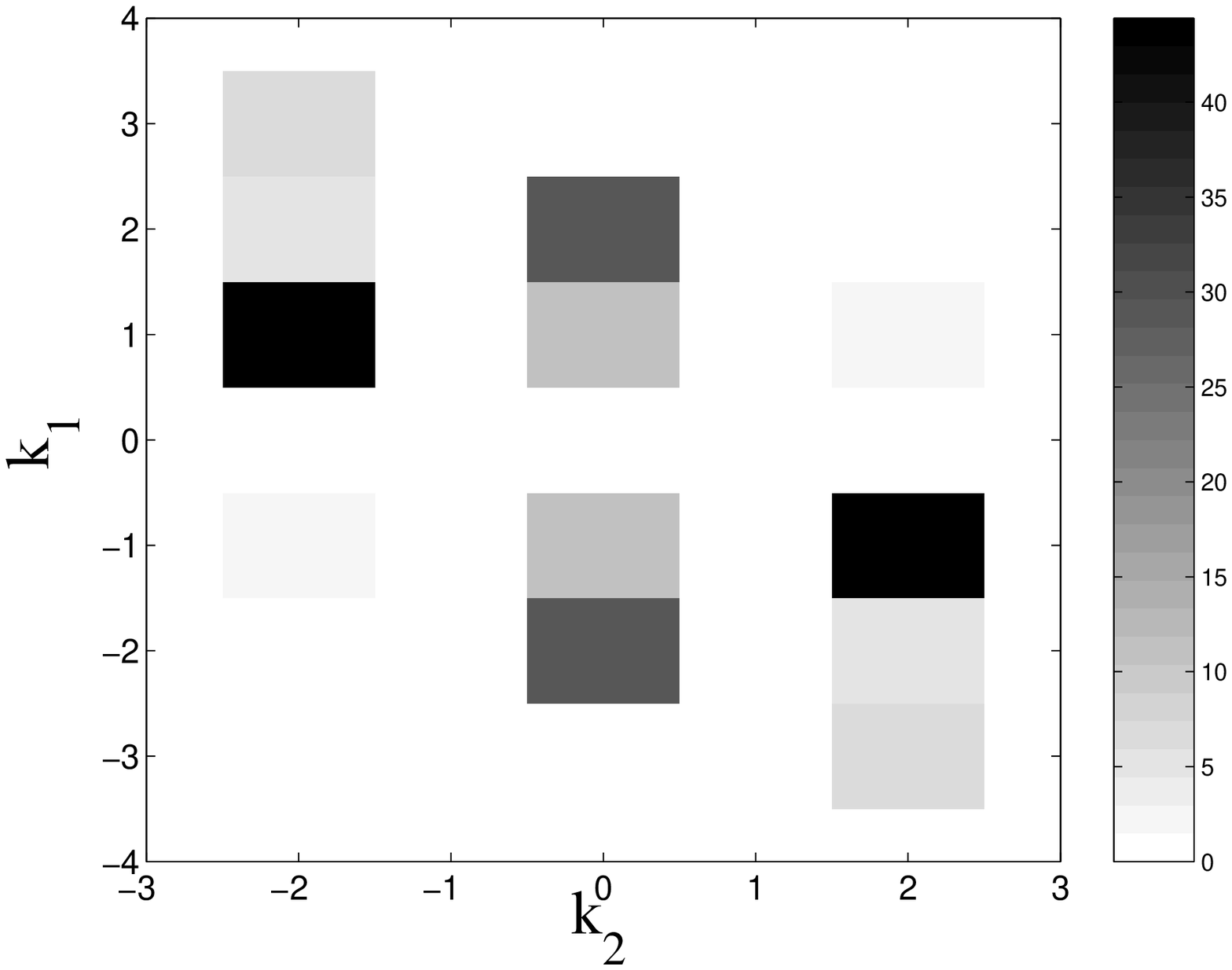}
 \mbox{{\bf (a)} }
 \end{minipage}
 \begin{minipage}[t]{8cm}
 \centering
 \includegraphics[width=6.cm,height=5.4cm]{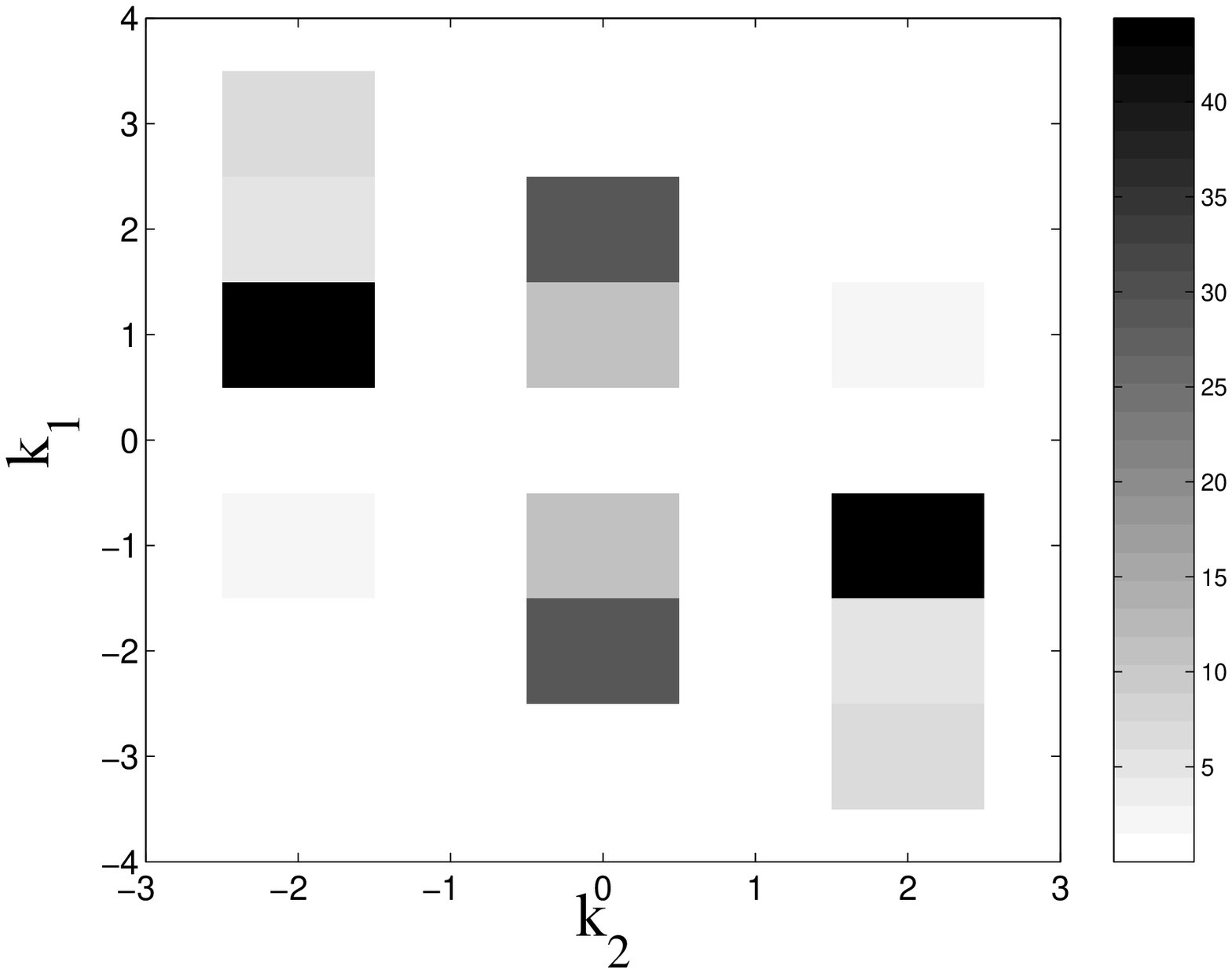}
 \mbox{{\bf (b)} }
 \end{minipage}
 \caption {\label{fig:fig3} Two-dimensional Fourier transforms of (a) $f$ given by Eq.~(\ref{eqnf}) and (b) $f_2$ given by Eq.~(\ref{eqnf2}) for $\lambda=0.03$ and $\omega_{0}=0.6750$.}
\end{figure*}

However, the control terms $f$ and $f_2$ given by
Eqs.~(\ref{eqnf}) and (\ref{eqnf2}) appear to have a much richer
Fourier spectrum. We have represented in Fig.~\ref{fig:fig3} their
two-dimensional Fourier transforms. They have an infinite number
of Fourier modes and therefore not practical for a numerical or
experimental realization. However, it is seen on
Fig.~\ref{fig:fig3} that only few Fourier coefficients contribute
significantly to the control terms. Therefore, it is feasible to
truncate them since the method has been shown to be
robust~\cite{Ciraolo1}. The tailored control term results in
general from a trade-off between the ability to control chaos and
restrictions on the desired shape for the specific problem at
hand.

In order to identify the main Fourier modes, we introduce a
parameter $A$ defined as
$$
A_{k_{1},k_{2}}\equiv\frac{|f_{k_{1},k_{2}}|}{|k_{1}\omega_{0}+k_{2}|},
$$
where $f_{k_{1},k_{2}}$ is the Fourier coefficient with wavevector
$(k_{1},k_{2})$ of $f$ or $f_2$. The dominant Fourier mode is
supposed to have maximal $A$. We notice that this definition
contains two effects~: First a dominant Fourier mode has to have a
significant amplitude, and second, its corresponding wavevector
has to be close to a resonance with the frequency vector of the
integrable motion (and hence close to a resonance).  For
$\omega_{0}=0.6750$ (which corresponds to $n_0=1$) and
$\lambda=0.03$, there is only one dominant Fourier mode in $f$ or
$f_2$ which has a frequency which is twice the microwave
frequency. The truncated control term is given by
\begin{equation}
\label{fa32} f_a(\theta,t)= f_{3,-2}\cos(3\theta-2t),
\end{equation}
where $f_{3,-2}\approx -9.772\times10^{-4}$ for control term $f$
given by Eq.~(\ref{eqnf}) and $f_{3,-2}\approx
-9.739\times10^{-4}$ for the approximate control term $f_2$ given
by Eq.~(\ref{eqnf2}). We notice that these two values are very
close. For this mode, we have $A_{3,-2}\approx 3.90\times10^{-2}$
which is more than ten times larger than the second largest one
$A_{1,-2}\approx 1.90\times10^{-3}$. We notice that the continued
fraction expansion of $\omega_0$ is $[0,1,2,\ldots]$. One best
approximant is $[0,1,2]=2/3$ which is the frequency of the mode of
$f_a$.

\begin{figure*}
 \begin{minipage}[t]{8cm}
 \centering
 \includegraphics[width=5.cm,height=5.cm]{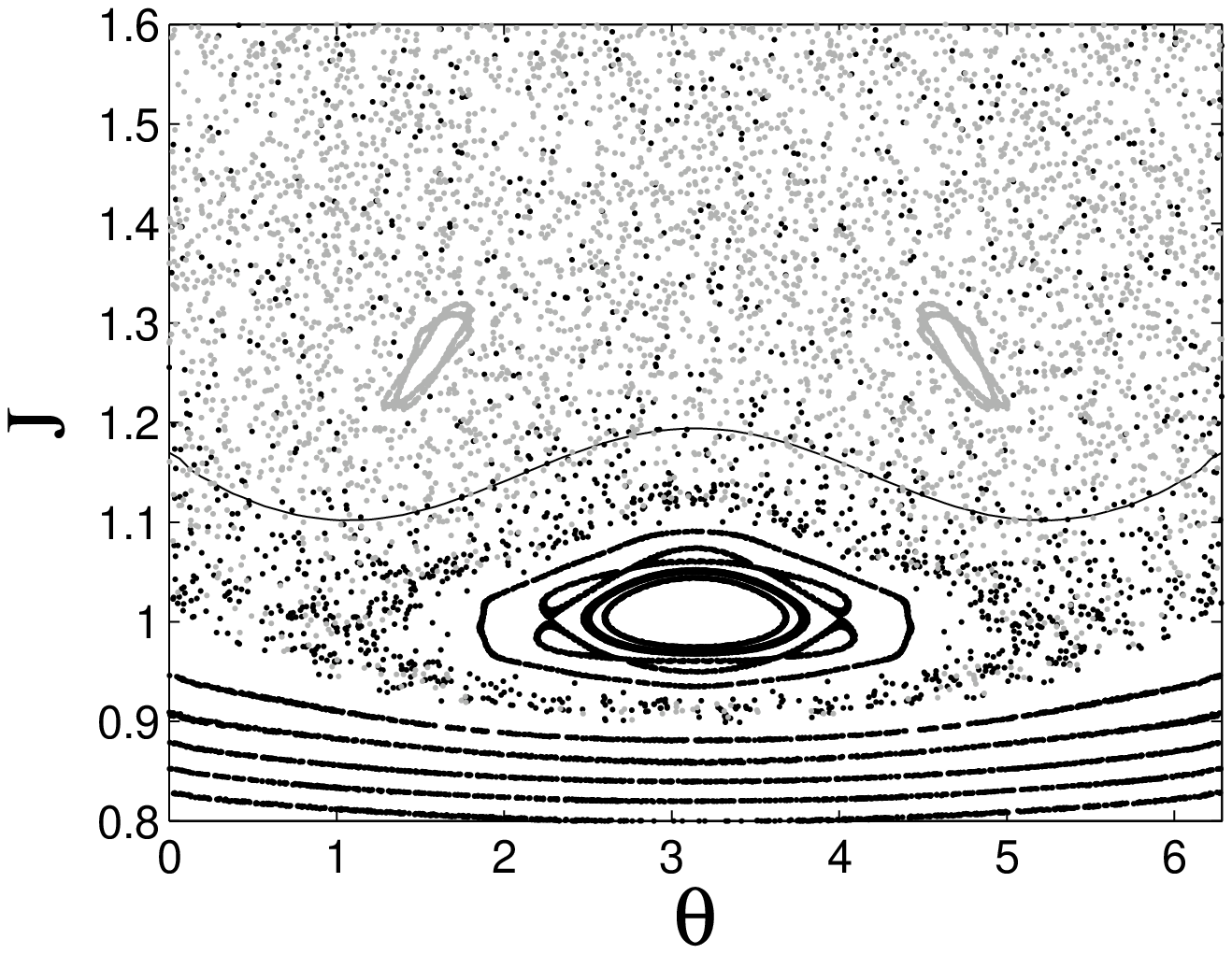}
 \mbox{{\bf (a)} }
 \end{minipage}
 \centering
 \begin{minipage}[t]{8cm}
 \includegraphics[width=5.cm,height=5.cm]{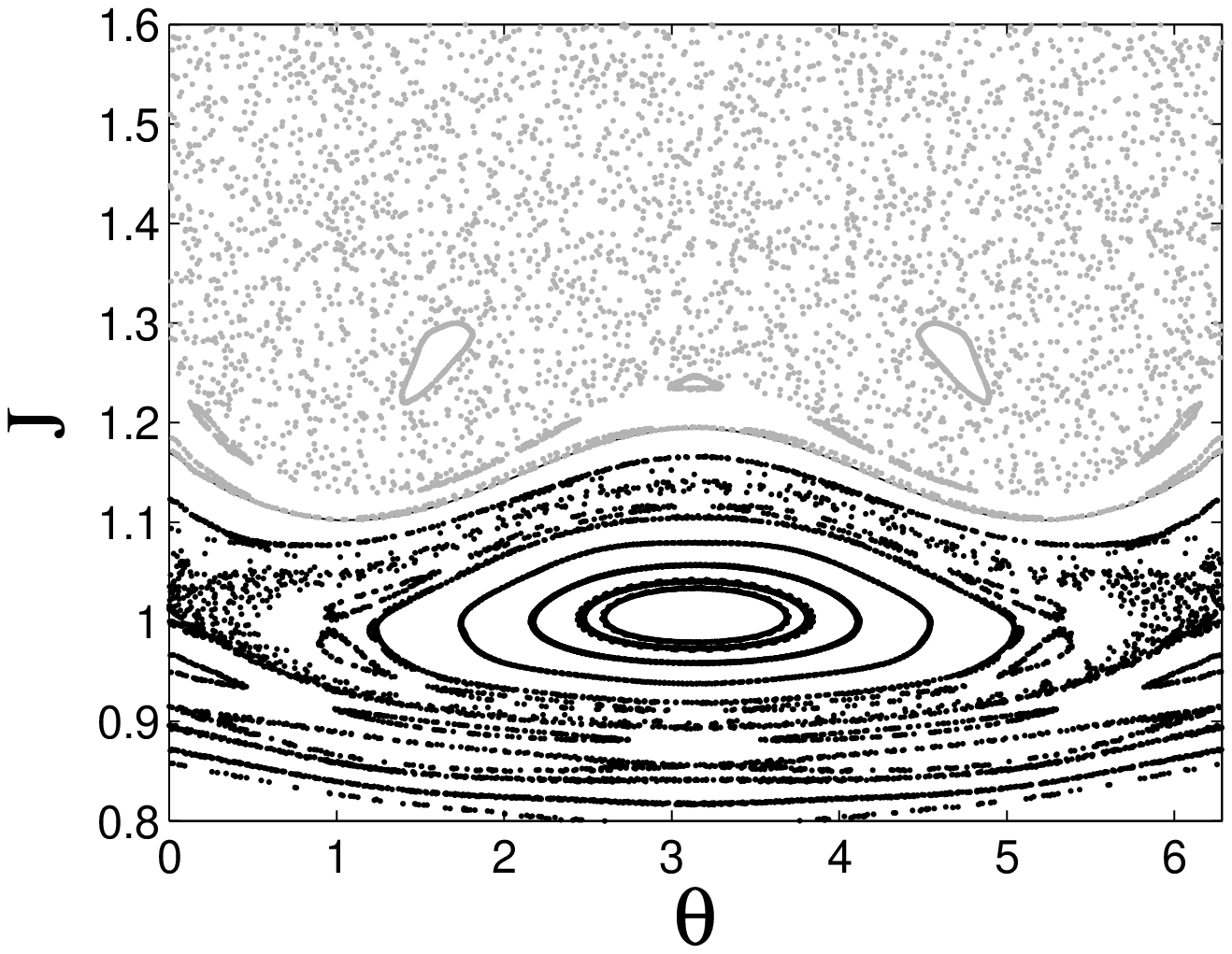}
 \mbox{{\bf (b)} }
 \end{minipage}
 \begin{minipage}[t]{8cm}
 \centering
 \includegraphics[width=5.cm,height=5.cm]{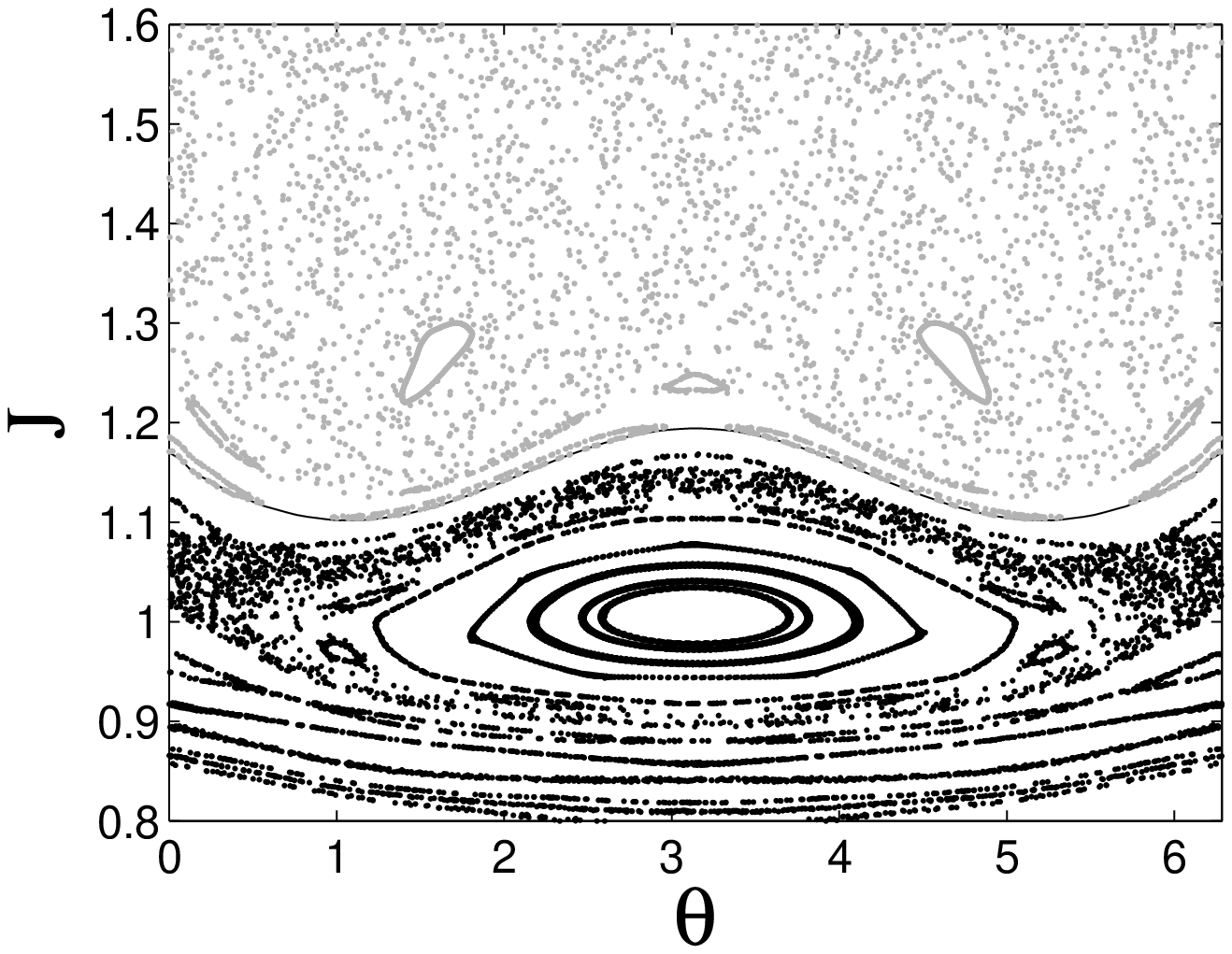}
 \mbox{{\bf (c)} }
 \end{minipage}
 \centering
 \begin{minipage}[t]{8cm}
 \includegraphics[width=5.cm,height=5.cm]{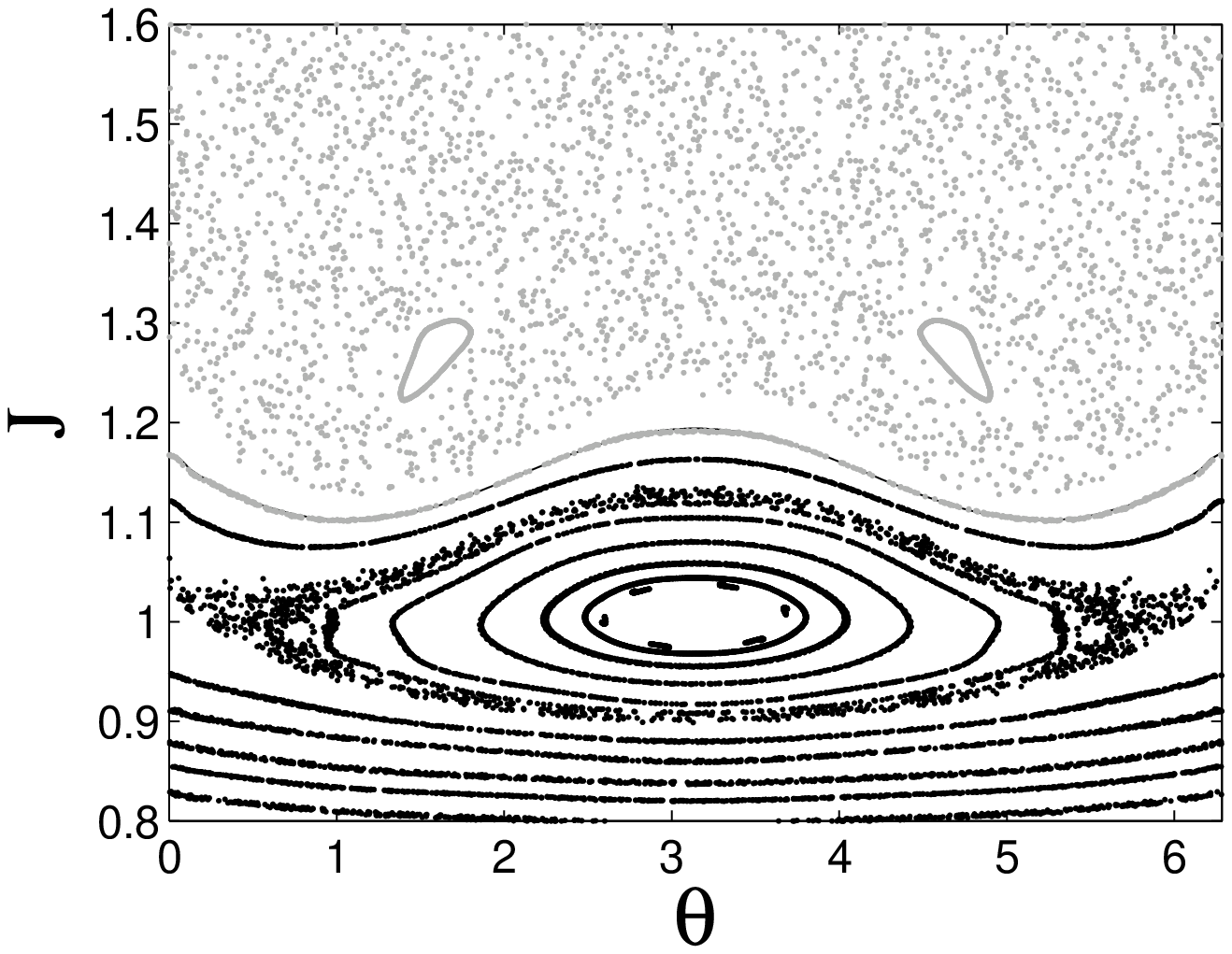}
 \mbox{{\bf (d)} }
 \end{minipage}
 \caption {\label{fig:fig4} Poincar\'e sections of $(a)$ The uncontrolled Hamiltonian $H$ given by Eq.~(\ref{HatomAA}),
 $(b)$ The controlled Hamiltonian $H+f$ where $f$ is given by Eq.~(\ref{eqnf}), $(c)$ The controlled
 Hamiltonian $H+f_2$ where $f_2$ is given by Eq.~(\ref{eqnf2}), and $(d)$
The controlled Hamiltonian $H+f_a$ where $f_a$ is given by
Eq.~(\ref{fa32}) for $\lambda=0.03$ and
 $\omega_{0}=0.6750$. The
 thin wary curve indicates the location where the
invariant torus is restored. The black dots are from trajectories
launched below this curve, and gray dots are from trajectories
launched above this curve. Note how they are interspersed in
$(a)$, as is expected of chaotic trajectories, and how the control
restricts their movements in phase space through the invariant
torus.}
\end{figure*}

\subsection{Poincar\'e sections}

In order to test the efficiency of the control terms to restore
invariant tori in phase space, we perform Poincar\'e sections of
$H+f$, $H+f_2$ and $H+f_a$ and compare them to the Poincar\'e
section of $H$ given by Eq.~(\ref{HatomAA}). Since all these
Hamiltonians are periodic in time with period $2\pi$, the natural
Poincar\'e section is a stroboscopic plot with period $2\pi$.

Figure~\ref{fig:fig4} depicts Poincar\'e sections of
Hamiltonian~(\ref{HatomAA}) in panel $(a)$, Hamiltonian $H+f$
where $f$ is given by Eq.~(\ref{eqnf}) in panel $(b)$, Hamiltonian
$H+f_2$ where $f_2$ is given by Eq.~(\ref{eqnf2}) in panel $(c)$
and Hamiltonian $H+f_{a}$ where $f_{a}$ is given by
Eq.~(\ref{fa32}) in panel $(d)$ for $\omega_{0}=0.6750$ and
$\lambda=0.03$. We notice that with the addition of the control
terms, an invariant torus has been restored which prevent the
diffusion from below to above the invariant torus. It is also
worth noticing that all of these control terms are efficient
although only $f$ is expected to be, indicating that the presence
of the control field $f_2$ contributes dominantly to a restoration
of invariant tori at specific locations such that higher order
resonances are eliminated which, in the Chirikov's approach
\cite{Chirikov}, leads to less chaos and hence less stochastic
ionization \cite{Cary,Chandre2} in our problem. It reinforces the
robustness of the method and allows one to tailor a control term
which is simpler to implement.

\begin{figure*}
 \centering
 \includegraphics[width=5.cm,height=5.cm]{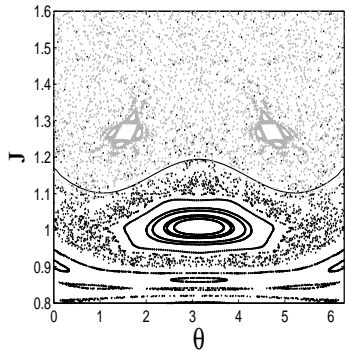}
 \caption {\label{fig:fig5}Poincar\'e sections of Hamiltonian~(\ref{eqnHwAA}) for
$\lambda=0.03$ and $\mu=0.0127$.}
\end{figure*}

\begin{figure*}
 \begin{minipage}[t]{8cm}
 \centering
 \includegraphics[width=6cm,height=5.4cm]{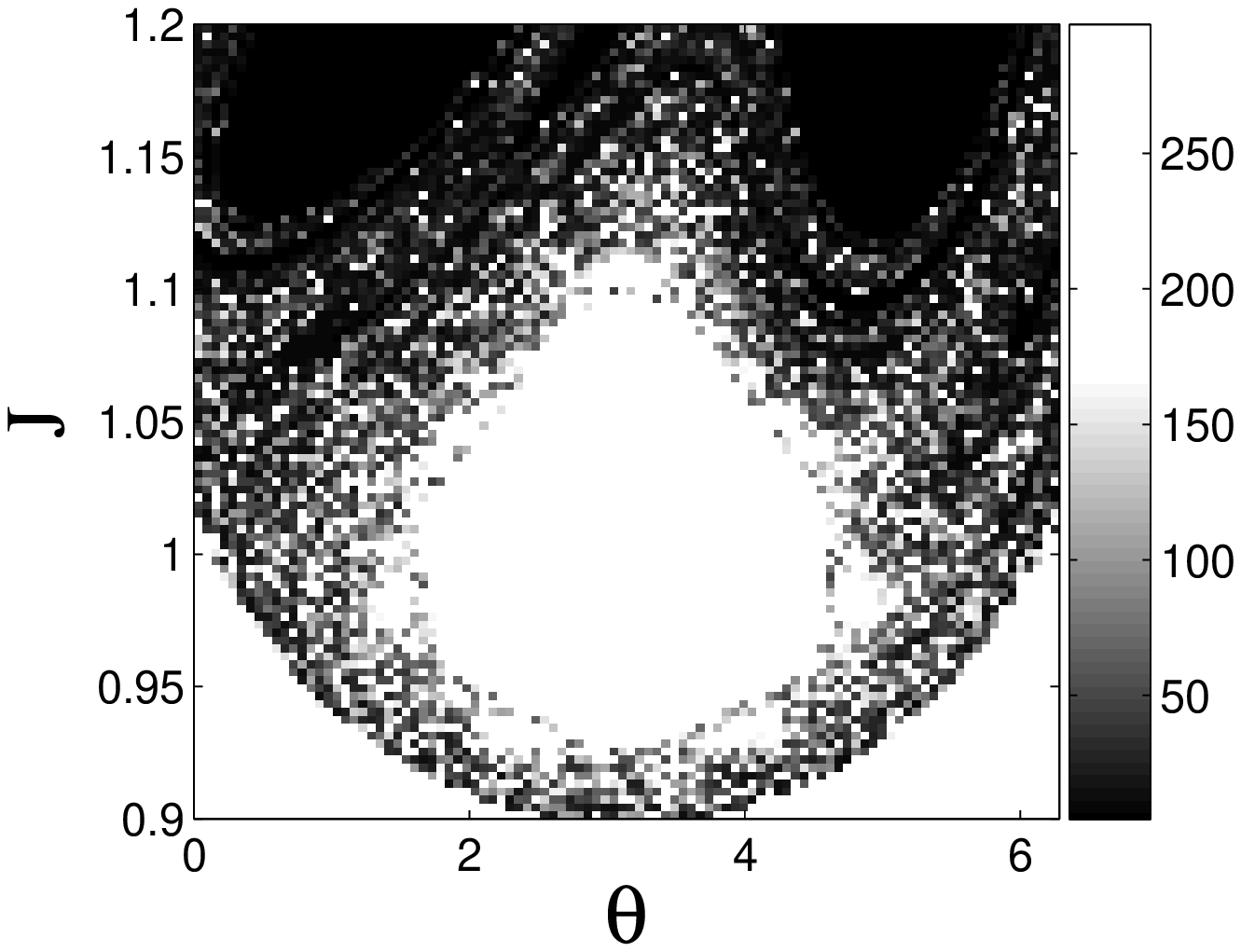}
 \mbox{{\bf (a)} }
 \end{minipage}
 \centering
 \begin{minipage}[t]{8cm}
 \includegraphics[width=6cm,height=5.4cm]{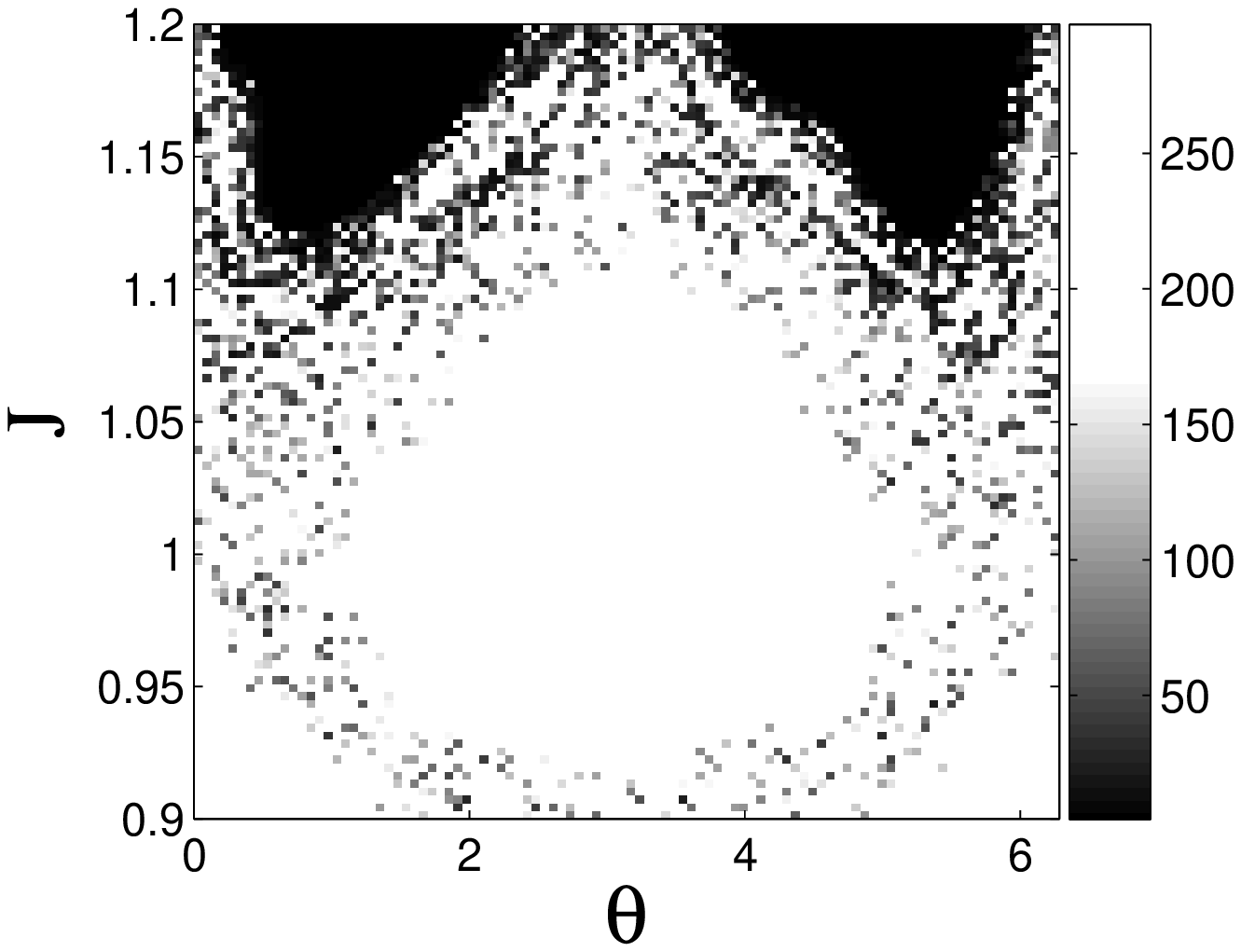}
 \mbox{{\bf (b)} }
 \end{minipage}
 \caption {\label{fig:fig6} Laminar plots of $(a)$ Hamiltonian~(\ref{HatomAA}) and $(b)$ Hamiltonian~(\ref{eqnHwAA}) for
$\lambda=0.03$ and $\mu=0.0127$. Cut-off time is $600\pi$ and
diffusion threshold is $J_{th}=1.30$.}
\end{figure*}

\begin{figure*}[!htb]
 \centering
 \includegraphics[width=5.6cm,height=5.6cm]{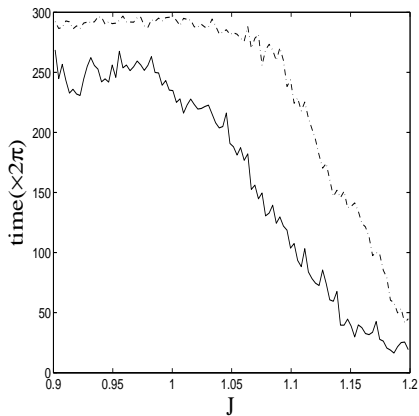}
 \caption {\label{fig:fig7}Mean diffusion time $\langle T_d\rangle$ versus initial action $J$ with
 amplitude of external field $\lambda=0.03$. Solid lines are for Hamiltonian~(\ref{HatomAA}). Dash-dotted lines are for Hamiltonian~(\ref{eqnHwAA}) for $\mu=0.0127$. Cut-off time is $600\pi$ and diffusion threshold is $J_{th}=1.30$. }
\end{figure*}

\subsection{A control term as an additional wave}

In Sec.~\ref{sec3A}, we show that the frequency of the control
wave should be twice the one of the initial wave
\footnote{Bichromatic microwave experiments are commonly used in
manipulating microwave ionization
\cite{Ko,Sirko1,Sirko2,PMKoch}.}. Therefore, a possible controlled
Hamiltonian is
\begin{equation}
H=\frac{p^{2}}{2}-\frac{1}{x}+\lambda x\cos t+\mu x\cos 2 t,
\end{equation}
which corresponds to a control terms $g(t)=\mu\cos 2t$ in
Eq.~(\ref{Hc}). In order to obtain the value of $\mu$, we use the
Fourier decomposition of the control term $f$ obtained previously.
First we map the controlled Hamiltonian into action-angle
variables~:
\begin{equation}
\label{eqnHwAA} H=-\frac{1}{2J^{2}}+2J^{2}\left(\lambda\cos t+\mu
\cos 2t\right) v(\theta, t),
\end{equation}
where $v$ is given by Eq.~(\ref{eqnv}). If we give a value of the
action $J_0$ where the invariant torus has to be restored, we have
seen that the dominant Fourier mode is proportional to
$\cos[(2n+1)\theta-2t]$ where $n$ is obtained using the continued
fraction expansion of $\omega_0=J_0^{-3}$. This mode is present in
Eq.~(\ref{eqnHwAA}) and has an amplitude given by $\mu J_0^2
a_{2n+1}$. If $f_{2n+1,-2}$ denotes the amplitude of the dominant
Fourier mode in Eq.~(\ref{eqnf}) for the values of the parameters
$\omega_0$ and $\lambda$, then the amplitude of the control field
is chosen to be
\begin{equation}
\label{eqnmu} \mu=\frac{f_{2n+1,-2}}{J_0^2 a_{2n+1}}.
\end{equation}
The real parameters taken for the external microwave field and
control field are flexible for a set of rescaled amplitude of
external field $\lambda=0.03$ and frequency $\omega=1$ as long as
they satisfy rescaling relationships.

Figure~\ref{fig:fig5} depicts the Poincar\'e section of the
controlled Hamiltonian~(\ref{eqnHwAA}) with $\lambda=0.03$ and
$\mu=0.0127$. Figure~\ref{fig:fig5} does not show the restoration
of an invariant torus as in the previous cases and hence, the
ionization reduction is not obvious. This comes from the fact that
the additional wave is quite far from the control
term~(\ref{eqnf}) due to additional resonances which break-up the
restored invariant torus.

However, the ionization process is still reduced, and this can be
seen by looking at laminar plots. Such plots are obtained by
looking at a grid of initial conditions and plotting the number of
iterations it takes the action to exceed a certain threshold.
Figure~\ref{fig:fig6} depicts the laminar plots for
Hamiltonian~(\ref{HatomAA}) and Hamiltonian~(\ref{eqnHwAA}) with
$\lambda=0.03$ and $\mu=0.0127$. The action threshold is chosen to
be $J_{th}=1.30$. The maximum integration time is $600\pi$. The
darker the region is the smaller time it takes to have $J\geq
J_{th}$.  It is expected that the laminar plots with more brighter
regions are cases where there is less ionization.

In order to compare the diffusion time of trajectories for
Hamiltonian~(\ref{HatomAA}) with that of the controlled
Hamiltonian~(\ref{eqnHwAA}), we have taken a set of \textsl{N}
initial angles evenly distributed in $[0,2\pi]$ for one initial
action $J$ and then computed the mean diffusion time for each $J$
in both controlled and uncontrolled cases~:
\begin{equation}
\langle T_{d}\rangle
(J)=\frac{1}{\textsl{N}}\sum_{i=1}^{\textsl{N}}T_{d}(J,\theta_i).
\end{equation}

Figure~\ref{fig:fig7} depicts the the curve of mean diffusion time
$\langle T_d\rangle$ versus initial action $J$. In the numerical
computation of $T_{d}(J,\theta_{i})$, the integration is performed
till the cut-off time $t=600\pi$. Therefore for some trajectories
the actual diffusion time is certainly above the cut-off time or
even goes to infinity. The double frequency control field also
works for the regime $1.14<J<1.20$ for the reason that the rebuilt
invariant torus is a curve which goes beyond
$J_{0}=\omega_{0}^{-1/3}=1.14$ for some areas.
Figure~\ref{fig:fig7} shows that the mean diffusion time for
controlled Hamiltonian~(\ref{eqnHwAA}) is significantly larger
than that for Hamiltonian~(\ref{HatomAA}) which clearly shows the
effect of the additional microwave field to reduce ionization.

\section{Conclusion}
\label{sec4}

In this paper we implemented a local control method on the
one-dimensional hydrogen atom in a linearly polarized (LP)
microwave field in order to reduce ionization. After simplifying
the originally complicated control function numerically, we
obtained an extremely simple control term which is in the same
form as the external LP microwave field but with smaller
amplitude. Adding the small control field to the perturbed
Hamiltonian leads to a reduction of ionization. We have done the
calculations in a regime where the quantum and classical agree,
and our classical computations show efficient suppression of
ionization. Preliminary results we obtained for controlling
ionization in higher dimensions show the promise of the local
control method for manipulating ionization in full-dimensional
Rydberg atoms and work in this direction is currently under way in
our center.

\begin{acknowledgments}
This research was supported by the US National Science Foundation.
CC acknowledges support from Euratom-CEA (contract
EUR~344-88-1~FUA~F).
\end{acknowledgments}


\begin{thebibliography}{99}


\bibitem{Gallagher9}
T. F. Gallagher, Rydberg Atoms (Cambridge University Press,
Cambridge, UK, 1994)

\bibitem{Bayfield}
J.E. Bayfield and P.M. Koch, Phys. Rev. Lett. \textbf{33}, 258
(1974)

\bibitem{Connerade}
J.-P. Connerade, Highly Excited Atoms (Cambridge University Press,
Cambridge, UK, 1998)

\bibitem{Koch}
P. M. Koch and K. A. H. van Leeuwen, Phys. Rep. \textbf{255}, 289
(1995)

\bibitem{Casati5}
G. Casati, B.V. Chirikov, I. Guarneri, D.L. Shepelyansky, Phys.
Rep. \textbf{154}, 77 (1987)

\bibitem{Casati}
G. Casati, I. Guarneri, and D.L. Shepelyansky, IEEE J. Quantum
Electron. \textbf{24}, 1420 (1988)

\bibitem{Jensen}
R. V. Jensen, Phys. Rep. \textbf{201}, 1 (1991)

\bibitem{Meerson}
B.I. Meerson, E.A. Oks and P.V. Sasorov, Pis'ma Zh. Eksp. Teor.
Fiz. \textbf{29}, 79 (1979) [JETP Lett. \textbf{29}, 72 (1979)]



\bibitem{Leopold3}
J.G. Leopold and I.C Percival, J. Phys. B \textbf{12}, 709 (1979)

\bibitem{Jensen1}
R. V. Jensen, Phys. Rev. A \textbf{30}, 386 (1984)

\bibitem{MacKay}
R. S. MacKay and J. D. Meiss, Phys. Rev. A \textbf{37}, 4702
(1988)

\bibitem{Farrelly}
D. Farrelly and T. Uzer, Phys. Rev. A, \textbf{38}, 5902 (1988)

\bibitem{Howard1}
J.E. Howard, Phys. Lett. A \textbf{156}, 286 (1991)

\bibitem{Howard2}
J.E. Howard, Phys. Rev. A \textbf{46}, 1 (1992)

\bibitem{Reichl}
L.E. Reichl, The Transition to Chaos in Conservative Classical
Systems: Quantum Manifestations (Springer-Verlag, New York, 1992)



\bibitem{Blumel9}
R. Bl\"umel and W. P. Reinhardt, Chaos in Atomic Physics
(Cambridge University Press, Cambridge, UK, 1997)

\bibitem{Buchleitner}
A. Buchleitner, D. Delande and J.-C. Gay, J. Opt. Soc. Am. B
\textbf{12}, 505 (1995)

\bibitem{Krug}
A. Krug and A. Buchleitner, Phys. Rev. A \textbf{72}, 061402(R)
(2005)

\bibitem{Berry}
M.V. Berry, Proc. R. Soc. Lond. A \textbf{413}, 183 (1987)

\bibitem{Leopold}
J.G. Leopold and D. Richards, J. Phys. B \textbf{18}, 3369 (1985)

\bibitem{Leopold1}
J.G. Leopold and D. Richards, J. Phys. B \textbf{22}, 131 (1989)

\bibitem{Blumel}
R. Bl\"umel and U. Smilansky, Phys. Scr. \textbf{40}, 386 (1989)

\bibitem{Delande}
D. Delande, Chaos in atomic and molecular physics, in: M. J.
Giannoni, A. Voros, J. Zinn-Justin(Eds), Chaos and Quantum
Physics, Les Houches, Session LII, 1989, Elsevier, Amsterdam,
1991, p.665

\bibitem{Perotti}
L. Perotti, Phys. Rev. A \textbf{73}, 053405 (2006)

\bibitem{Ko}
L. Ko, M. W. Noel, J. Lambert and T. F. Gallagher, J. Phys. B
\textbf{32}, 3469 (1999)

\bibitem{Sirko1}
L. Sirko, S. A. Zelazny, and P. M. Koch, Phy. Rev. Lett.
\textbf{87}, 043002 (2001)

\bibitem{Sirko2}
L. Sirko and P. M. Koch, Phy. Rev. Lett. \textbf{89}, 274101
(2002)

\bibitem{PMKoch}
P. M. Koch, S. A. Zelazny and L. Sirko, J. Phys. B: At. Mol. Opt.
Phys. \textbf{36}, 4755 (2003)

\bibitem{Gallagher}
H. Maeda and T. F. Gallagher, Phy. Rev. Lett. \textbf{93}, 193002
(2004)

\bibitem{Maeda9}
H. Maeda, J. H. Gurian, D. V. L. Norum and T. F. Gallagher, Phys.
Rev. Lett. \textbf{96}, 073002 (2006)

\bibitem{Prosen}
T. Prosen and D. L. Shepelyansky, Eur. Phys. J. B \textbf{46}, 515
(2005)

\bibitem{Chandre2}
C. Chandre, M. Vittot, G. Ciraolo, Ph. Ghendrih and R. Lima, Nucl.
Fusion \textbf{46}, 33 (2006)

\bibitem{Chirikov}
B.V. Chirikov, Phys. Rep. \textbf{52}, 263 (1979)

\bibitem{Ciraolo1}
G. Ciraolo, C. Chandre, R. Lima, M. Vittot, M. Pettini, C.
Figarella and  P. Ghendrih, J. Phys. A: Math. Gen. \textbf{37},
3589 (2004)

\bibitem{Casati9}
G. Casati, B. V. Chirikov, D. L. Shepelyansky, and I. Guarneri,
Phys. Rev. Lett. \textbf{57}, 823 (1986)

\bibitem{Galvez}
E. J. Galvez, B. E. Sauer, L. Moorman, P. M. Koch, and D.
Richards, Phys. Rev. Lett. \textbf{61}, 2011 (1988)

\bibitem{Cary}
J. R. Cary and J. D. Hanson, Phys. Fluids \textbf{29}, 2464 (1986)




\end{thebibliography}
\end{document}